\documentclass{aa}
\usepackage{graphics}

\newcommand{\ec}{\ensuremath{E_\mathrm{c}} }
\newcommand{\al}{\ensuremath{\Gamma} }
\newcommand{\chisq}{\ensuremath{\chi^{2}} }
\newcommand{\be}{\ensuremath{\beta} }
\newcommand{\nh}{\ensuremath{N_\mathrm{H}} }
\newcommand{\tetref}{\ensuremath{\theta_\mathrm{0}} }
\begin{document}

   \thesaurus{11     
              (02.18.7;  
               13.25.2;  
               11.19.1;  
               11.09.1 \object{IC 4329a}  
               11.09.1 \object{MGC 8-11-11};  
               11.09.1 \object{NGC 4151})} 
   \title{Anisotropic Illumination in AGNs} 
   \subtitle{The reflected component.\\ Comparison to hard X-ray spectra from Seyfert Galaxies}

   \author{J. Malzac
          \inst{1}
          \and
          E. Jourdain\inst{1}
	 \and
         P.O. Petrucci\inst{2}
         \and
         G.Henri\inst{2}
	}
         \offprints{J. Malzac}
	\mail{malzac@cesr.cnes.fr}

   \institute{C.E.S.R. (CNRS/UPS), 9, av. du Colonel Roche, B.P. 4346, 31028 Toulouse Cedex 4, France\\
        \and
             Laboratoire d'Astrophysique, Observatoire de Grenoble, B.P. 53X, F38041 Grenoble Cedex, France\\
        }

   \date{Received ??; accepted ??}

   \maketitle

   \begin{abstract}

     We calculate the reflection component predicted by the anisotropic illumination model of Henri \& Petrucci \cite{henri}. This component appears to be more important than for isotropic models. The overall X/$\gamma$ spectrum is found to be strongly angle-dependent. When the accretion disc is seen with a nearly edge-on orientation the reflection hump is weak, while a face-on viewing angle leads to a proeminent reflection hump with  an equivalent reflection coefficient $R\sim50$. Such reflection dominated Seyfert 1s galaxies are not observed. By fitting observed X/$\gamma$ spectra, we derive inclination angle $\theta\sim70\degr$ for \object{MCG 8-11-11} and $\theta\sim80\degr$ for \object{IC 4329a} and \object{NGC 4151}. Although the model succeed in reproducing individual observed spectra, it requires all the Seyfert 1s observed in the X-ray band to be seen with large inclination angles. Such a situation is highly improbable. On the other hand, we show that the ionisation of a fair part of the reflecting disk could represent an interesting improvement of the model, consistent with the data and relaxing the constraints on the high energy cut-off in Seyfert galaxies.

   \end{abstract}
 

\section{Introduction}

Although we know now that the high energy spectrum of AGNs is driven by  accretion processes in the vicinity of a supermassive black hole, the detailed nature and geometry of the nuclear emitting region are still unclear.

The radio quiet AGNs have first been divided in two spectral classes based on optical classification (Khali\-chi\-kyan \& Weedman \cite{khalichi}).The Seyfert 1s show broad (up to 10000 km/s) and narrow (up to 1000 km/s) emission lines while the seyfert 2s show only the narrow ones. The broad lines originate probably from hot photoionized clouds\\within 1 parsec of the central source, while the narrow lines originate in  moving clouds at kilo parsec distances. The most popular models to explain these differences are the unified models (see Antonucci \cite{antonucci} for a review).  In this scheme Seyfert 1s and 2s have intrinsically the same nucleus, the differences are simply due to an inclination effect. An optically thick torus of dust and molecular gas surrounds the nucleus and the broad line region (BLR), obscuring our view of the BLR when our line of sight lies close to the torus plane. In this case the galaxy appears as a Seyfert 2 since only the NRL are visible. The BLR and the nucleus are visible only when we view from within a cone aligned with the polar axis of the torus. The galaxy then appears as Seyfert 1. The most extreme hypothesis, called the straw's person model, assumes that the thickness and the opening angle of the torus are the same in all Seyfert galaxies. This model received strong supports from observations, but there are now some hints that the reality is more complex lying somewhere between the straw's person model and the hypothesis that Seyfert 1s and 2s are intrinsically different objects.

Since the high energy photons are less altered by the environment, and
because of their short time variability, the observed X and $\gamma$ spectra provide information on the inner regions and the primary source.
 They are well described by an intrinsic power law with photon index $\Gamma\sim~1.9$ (Pounds et al. \cite{pounds90}; Nandra \& Pounds \cite{nandra94}) with a cut-off at a few hundred keV (Jourdain et al. \cite{jourdain}; Johnson et al. \cite{johnsonco}). This powerlaw is interpreted as the result of comptonisation of soft (UV) photons by high energy electrons or pairs. Ginga observations have shown the presence of secondary components surimposed on the powerlaw. Among them a reflected component arising from Compton reflection on cold matter (White \& Lightman \cite{white}) and a neutral \ion{Fe}{} $\mathrm{K_{\alpha}}$ line. These components suggest the presence of a cold thick gas in the region where most of the power is released.

This is corroborated by the presence in many sources of a strong UV bump interpreted as a thermally radiating medium. This UV emission was thought to arise from internal dissipation in an accretion disc until variability studies of \object{NGC 4151} and \object{NGC 5548} (Clavel et al. \cite{clavel}; Warwick et al. \cite{warwick}; Perola et al. \cite{perola}) showed a correlated UV and X variability which is inconsistent with the standard accretion disc model. The UV and optical continua are more likely due to the  reprocessing of the high energy radiation.

All these elements have led to consider various geometries where an X-ray  source is located above a cold accretion disc and reilluminates it, inprinting the reflection features on the observed spectrum (Zdziarski et al. \cite{zdziarski90}; Haardt et Maraschi \cite{H&M91}, \cite{H&M93}). In these models the X-ray source is isotropic, half of the X-rays impinges on the disc and is reprocessed. This produces a black body UV luminosity of about the same magnitude as the hard Comptonized luminosity. 
This is contrary to observations which often show a UV luminosity several times greater than X-ray luminosity (Walter et Fink \cite{walter}). Furthermore, some observations of the \ion{Fe}{} line with a large equivalent width ($EW > 200$ eV, Nandra et al. \cite{nandra97}) seem to require more impinging radiation. Ghisellini et al. (\cite{ghisellini}) pointed out that the anisotropy of both the UV radiation field and the IC process should lead to more X-ray photons scattered backward than upward.

 However, the discovery by OSSE and SIGMA of an ubiquitous cut-off around a few hundred
keV has focused the attention on thermal models, stimulating detailed
calculations of the coupling between the hot source and the cold disc.
 The most popular model, favored by these calculations, is the patchy corona model (Haardt et al. \cite{haardtpc}; Stern et al. \cite{sternpc}). In this scheme the emitting region is modeled by a small number ($\sim 10$) of active regions on the surface of the accretion disc. This model is in good agreement with the data. However, it requires
internal dissipation in the disc to explain large UV to X ratios, which
is again difficult to reconcile with simultaneous optical-UV variability.
It seems also to have difficulties to explain photon index softer than 2, as
observed in some Seyfert 2s (Smith \& Done, \cite{smith}).

As an alternative,
Henri et Petrucci (\cite{henri}, hereafter HP97) proposed a model where the hard X-rays are emitted by  a unique non-thermal optically thin point source. This source is located above an accretion disc and illuminates it. Such a source could be physically realized by a strong shock terminating an aborted jet (Petrucci et al., in preparation). The seed photons are provided by the thermal emission of the disc. This disc is represented by an infinite slab which radiates only the reprocessed energy from the hot source whithout internal dissipation. The anisotropy of the soft photon  field influences the Compton process making the X/$\gamma$ emission strongly anisotropic. HP97 calculated in a self-consistent way the angular distribution of the high energy radiation and the disc temperature profile. This anistropic illumination model (AIM) was found to give a good explanation of the observed distribution of UV/X luminosity ratios in Seyfert 1s. The high energy cut-off can be obtained by an appropriate shape of the particle distribution that can be naturally produced by
reacceleration associated with pair production (Done et al. \cite{done90}, Henri \& Pelletier \cite{henri91}).
        
In this paper we investigate further this anisotropic illumination scenario.
We calculated the reflection component expected from this geometry (\emph{Sect.~2}). The strong anisotropy of the X-ray source leads to an overall spectrum which is highly angle dependent. For nearly pole-on viewing angles, the reflected component dominates the hard X-ray spectrum. In \emph{Sect.~3}, we compare our results with observed spectra of  \object{IC 4329a}, \object{MCG 8-11-11} and \object{NGC 4151} . The small amount of reflection in these objects requires nearly edge-on inclination angles. To date, no reflection dominated Seyfert 1 has been detected. It means
that \emph{in the framework of the present model all the X/$\gamma$
observed Seyfert 1s are seen with large inclinations}.
 This  point and possible implications for Seyfert galaxies are discussed in \emph{Sect.~4}.


\section{The reflection component}


\subsection{Computional Method and Assumptions}
 
The calculation of the reflected component of the proposed model is a bit peculiar, since the incident flux is angle dependent.
The flux from the primary X-ray source, radiated in a direction defined by an angle $\theta$ with the normal to the disc  was given in HP97 (Eq.~78) with a form similar to:
\begin{eqnarray}
 F_{\mathrm{dir}}\,(\,E\,,\,\mu) \propto F_{\eta\chi}(\mu)A(\mu)^{\Gamma-2} E^{1-\Gamma}\:\exp{-\sqrt\frac{E}{E_{\mathrm{c}}\,A(\mu)}}
\end{eqnarray}
where $\mu=\cos\theta$, $\Gamma$ is the photon index and $E_{\mathrm{c}}$ the cut-off energy.\\
The anisotropy coefficients are given by:
\begin{eqnarray}
F{\eta\chi}(\mu)&=& 2.30043\,-\,3.29152\,\mu\,+\,1.09871\,\mu^{2},\\
A(\mu)&=& \frac{F_{\eta\chi}(\mu)}{2.\,-\,1.64576\,\mu}.
\end{eqnarray}
Note that the effective cut-off energy $E_{\mathrm{c}}A(\mu)$ is angle dependent.\\
This form of $F_{\mathrm{dir}}$ differs from the classical power law plus exponential cut-off by the 1/2 exponent under the $\exp$ function. This exponent originates from the lepton energy distribution in the hot source. In HP97 (eq.~71), this distribution function has been \emph{arbitrary} fixed to be~:

\begin{eqnarray}
n(\gamma)=N_{0}\:\gamma^{-s}\exp\left(-\frac{\gamma}{\gamma_{0}}\right)
\end{eqnarray}
Actually, we should calculate it self consistently taking into account all radiative and hydrodynamic processes in the shock which leads to the formation of the hot source. This work is currently in progress and it seems that in the allowed range of parameters there are few constraints on the shape of the 
electron distribution. For example, we could choose the more general distribution :

\begin{eqnarray}
N(\gamma)=N_{0}\gamma^{-s}\exp\left[-\left(\frac{\gamma}{\gamma_{0}}\right)^{2\beta}\right]
\end{eqnarray}
\\
Which leads to a photon flux:

\begin{eqnarray}
F_{\mathrm{dir}}(E,\mu)\propto F_{\eta\chi}(\mu)A(\mu)^{\Gamma-2}E^{1-\Gamma}\nonumber\\
&\exp\left[-\left(\frac{E}{E_{\mathrm{c}}A(\mu)}\right)^{\beta}\right]
\label{fdir}
\end{eqnarray}
In this paper, we will use this last expression. Indeed, the model does not constrain the detailed shape of the primary spectrum but only the integrated luminosities.

For directions $\mu < 0$ , the radiation impinges on the disc. A fair part of the luminosity is absorbed contributing to the disc heating, while a fraction is Compton reflected leading to the formation of a secondary component $F_{\mathrm{ref}}$.
As for the isotropic case the strength and shape of reflected spectrum depend on the viewing angle (Magdziarz \& Zdziarski \cite{magdziarz}), but in a stronger manner. 
We  performed calculations of $F_{\mathrm{ref}}(\,E\,,\,\mu\,)$ using a Monte-Carlo code which enables a treatment of these anisotropy effects (See Poutanen et al. \cite{poutanen} for a semi-analytic treatment of angle dependent illumination). Our code  takes into account both Compton diffusions in cold matter, photoabsorption and iron fluorescence. The disc is supposed to be optically thick, we fixed its column density at $10^{26} \mathrm{cm^{-2}}$. The photoabsorption opacities are those from Morrisson \& Mc Cammon (\cite{morrison}) which assume neutral matter with standard abundances. This code was tested, for the iso\-tro\-pic case, against the results of Magdziarz \& Zdziarski \cite{magdziarz} and found in good agreement.
  
Assuming neutral matter means that the source height is supposed to be large enough for the impinging X-ray flux not to ionize the disc significantly.
This hypothesis is supported by the observations of the 6.4~keV \ion{Fe}{} line (Pound et al. \cite{pounds90}; Nandra \& Pounds \cite{nandra94}).
An ionized reflector would increase the reflected continuum below 10 keV where its contribution to the overall spectrum is usually very weak. 
 
In Petrucci \& Henri (\cite{petrucci}), general relativity effects were studied and found to have negligible influence on the primary and reprocessed spectrum for source heights greater than 25 $R_{s}$ ($5 \, 10^{12}$ to $5 \, 10^{14}$ cm for typical masses).  In this limit indeed, the characteristic size of the disc area where most of the source power is radiated, is much larger than the gravitational radius. Since our assumption of neutral matter requires a large source height, we assume that general relativity effects on the reflection component are negligible and use an Euclidean metric. Also we did not attempt to take into account the Doppler effect from the disc
rotation. These effects would affect essentially the \ion{Fe}{} line shape. 

The reflected luminosity does not contribute to the disc heating. This could
modify the temperature disc profile derived in HP97. Actually the disc albedo
is small and  we can neglect the incidence of reflection on the disc-source energy balance. We checked a posteriori that the total reflected luminosity is only about 10 $\%$ of the total incident luminosity.
 Also minor modifications of the primary spectrum arising from reflected photons going back to the hot source have been neglected.

With all these approximations the high energy spectrum shape is independent of the source height above the infinite disc.    

\begin{figure}
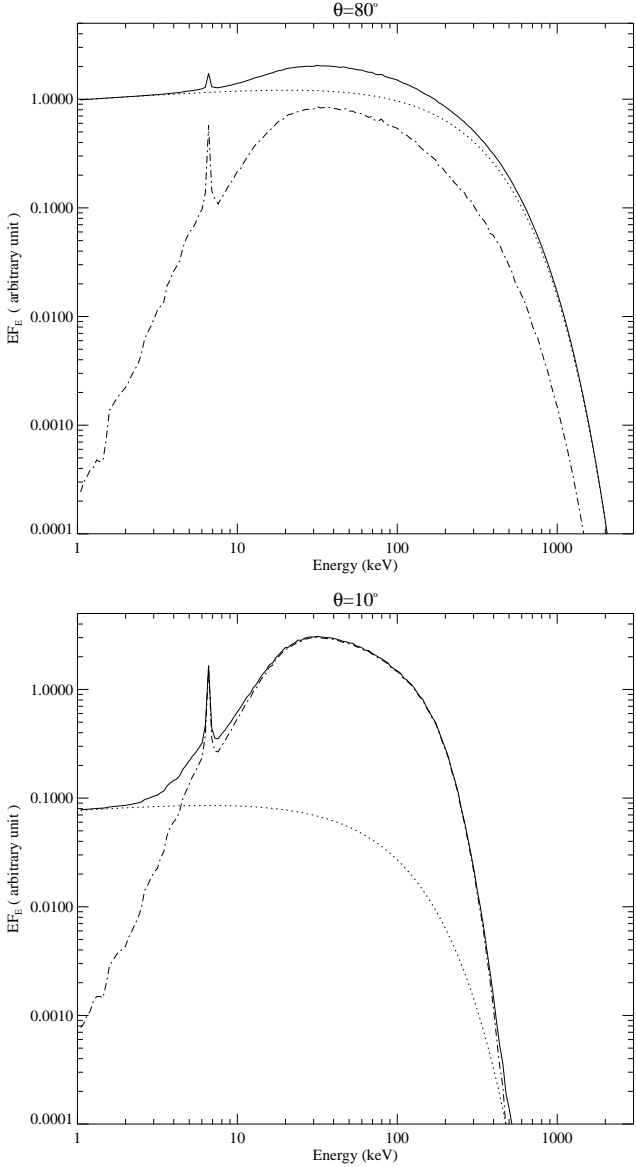

 \resizebox{\hsize}{!}{\includegraphics{s80_7524.f1}}
 \resizebox{\hsize}{!}{\includegraphics{s10_7524.f1}}
 \caption{Calculated  spectra for $\theta=10\degr$ (bottom) and $\theta=80\degr$ (top) with $\ec=200$ keV $\al=1.9$ and $\beta=1.$ The dotted lines show the primary spectrum, the dashed lines the reflected component.}   
 \label{calspec}
\end{figure}

\subsection{Results}

The main result from our calculations is that the overall spectrum is strongly dependent on the viewing angle $\theta$, defined as the angle between the line of sight and the normal to the disc. Fig.~\ref{calspec} shows two spectra obtained for edge-on and face-on inclinations. The hard X-ray source is highly anisotropic, the downward to upward emitted power ratio  is about 7, leading to a reflection component high\-er than for isotropic models. Furthermore, this anisotropy leads to an increasing apparent luminosity of the primary emission with increasing $\theta$ (see HP97). On the other hand the reflected flux is roughly proportional to the projected disc area seen by the observer which decreases as $\mu=\cos\theta$. So, \emph{the reflection component dominates 
the high energy spectrum for low inclinations}.

The amount of reflection in the overall spectrum can be quantified by the reflection coefficient $R$. This empirical parameter is used to account for anisotropy and geometry uncertainties when comparing an isotropic reflection model with data. $R$ is usually defined as the normalisation ratio between an observed reflected component to the calculated reflection obtained for an isotropic source illuminating an infinite slab seen with a given reference inclination angle \tetref.

 The choice of \tetref is important since even in the iso\-tro\-pic case the reflected spectrum shape and normalisation are angle dependent. Usually \tetref is chosen to be $60\degr$ because for this angle the spectrum is very  close to  the angle averaged spectrum, some  authors prefer $\tetref=0$ since Seyfert 1s are expected to be seen pole on.

 Here we estimate the reflection coefficient predicted by the AIM. Our  $R(\mu)$ is the ratio of the number of reflected photons predicted by the model for a given $\mu$, to the photon number obtained if the source was isotropic --
and emitting the spectrum predicted by our model for an inclination $\mu$ --  and observed with an inclination $\mu_{0}=\cos\tetref$:

\begin{eqnarray}
R(\mu)=\frac{\int{E^{-1}F_\mathrm{ref}(E,\mu)\,dE}}{\int{E^{-1}F_\mathrm{ref}^{\mathrm{iso},\mu}(E,\mu_{0})\,dE}}
\end{eqnarray}

\begin{figure}
 \resizebox{\hsize}{!}{\includegraphics{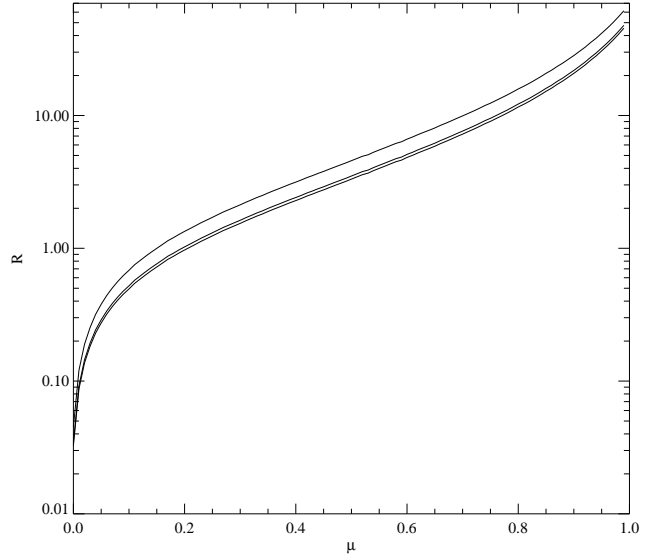}}
 \caption{R as a function of $\mu$ for three reference angles. $\tetref=0\degr$, $30\degr$ , $60\degr$ from bottom to top. The parameters of the primary emission are $\ec=200$ keV, $\al=1.9$, $\beta=1$.}
 \label{rdemu} 
\end{figure}

Fig.~\ref{rdemu} shows $R$ as a function of the inclination angle. 
$R$, increasing quickly with $\mu$, is of order of unity for angles around 80-75 degrees and become important for smaller angles reaching values of about 50 for small inclinations.

As the observed R in Seyfert 1 galaxies is of order unity, we can already see that the anisotropic illumination model predicts an important inclination angle for those Seyfert 1s that have been well studied in hard X-rays.
In the next section we test our model against the high energy spectra of \object{IC 4329a}, \object{MCG 8-11-11} and \object{NGC 4151} and derive inclination angles for these objects.

\section{Comparisons with observations}

\subsection{Data}

In order to test the anisotropic illumination model and its predicted reflection component, we need spectral data in the 1--500  keV range.
Seyfert 1 galaxies are faint at high energy, OSSE detected only 12 of them with a good photon statistic (Johnson et al. \cite{johnson}).
If we discard those which were not observed simultaneously by a lower energy instrument their number falls to only 3 objects for which data are available, namely 
\object{IC 4329a}, \object{MCG 8-11-11} and \object{NGC 4151}.

For \object{IC 4329a} we used the ROSAT, GINGA and OSSE data from Madejski et al. (\cite{madejski}, Fig. 4b model $d$). In our fits we ignored pha bins below 1 keV.
For \object{MCG 8-11-11} we used the ASCA/OSSE joint spectrum  described in Grandi et al.(\cite{grandi}).
For \object{NGC 4151} we used the data from Zdziarski et al. (\cite{zdziarski}). The original data are a composite spectrum based on nearly simultaneous observations by ROSAT, GINGA and OSSE in june 1991. \object{NGC 4151} presents a strong excess below 1.5 keV whose nature is unclear, arising probably from extended emission in the galaxy (Pounds et al \cite{pounds86}), and a strong complex absorption structure. The analysis of these components is out of the scope of the present work. So, we did not use the ROSAT data in our fit and considered GINGA data only above 2 keV.

\subsection{Fitted model}

We fit these data with a $3$ component model which has the following form:
\begin{eqnarray}
F\,(\,E\, , \,\mu\,) = \left[ \,F_{\mathrm{dir}}(\,E\,,\,\mu\,) + F_{\mathrm{ref}}(\,E\,,\,\mu\,)\,\right] e^{-\sigma\,(\,E\,)\,N_\mathrm{H}}
\end{eqnarray}

The first component is the direct flux $F_{\mathrm{dir}}$ defined by Eq.~\ref{fdir} which depends on parameters $\mu$, $E_\mathrm{c}$, $\Gamma$  and $\beta$.

The second component $F_\mathrm{ref}(E,\mu)$ is the reflected component.
This  component depends indirectly on parameters $E_\mathrm{c}$, $\Gamma$  and $\beta$ via the incident X-ray spectrum. The $\mu$ parameter affects weakly its shape, but mainly its relative intensity to the direct flux. 

As an attempt to account for the low energy absorption we add a simple neutral
absorption model with an hydrogen equivalent column density $N_\mathrm{h}$.
In order to have a reduced set of free parameters we did not attempt to model the data with a more complex absorber (ionized or partial covering). As we fit the data only above 1 keV we expect it does not influence the fit.
    
So, the fitted model includes 5 parameters $E_\mathrm{c}$, $\Gamma$, $\mu$, $N_\mathrm{h}$ and $\beta$ plus an overall normalisation.

The fits were performed using XSPEC V10. The computation of the reflected spectrum by Monte Carlo method is too time consuming to be implemented in the fit procedure. So, we calculated once for all the ``response matrix'' of the disc. This matrix 
(similar to those described in Poutanen et al. \cite{poutanen} ) gives the reflected spectrum in a given direction, for a given incident flux at fixed incidence angle and energy. An XSPEC implemented subroutine computes quickly the reflected component by interpolating and summing the matrix elements.

Quoted errors are the $90\%$ confidence range ($\Delta\chisq=2.706$) as computed by XSPEC.

\subsection{Results}



\begin{figure}

\resizebox{\hsize}{!}{\includegraphics{ic7524.f3}}
\end{figure}
\begin{figure}
\resizebox{\hsize}{!}{\includegraphics{mcg7524.f3}}
\end{figure}
\begin{figure}
 \resizebox{\hsize}{!}{\includegraphics{ngc7524.f3}}
\caption{Fits with model b. The dotted lines are the reflection components, dashed line are the absorbed primary spectra, dot-dashed lines are the unabsorbed continua, solid lines are the sum of the three components}   
 \label{data}
\end{figure}

We fitted the data with both the $\beta$ parameter fixed to 1 (models $a/c$) and let as a free parameter (models $b/d$). To estimate the influence of our predicted \ion{Fe} K$\alpha$ line on the results, we then reprocessed the fits ignoring the pha channels in the 6.-7. range (models $c$ and $d$).
The results are shown in table~\ref{results}. Fig.~\ref{data} shows the data and the unfolded model spectra from model $b$.




The best fit parameters values of \ec, \al  and \nh  derived  with $\be=1$ (model $a$) are very similar to those found in the previous quoted works. We found values of $\mu$ corresponding to~:
$\theta\sim81\degr$ for \object{IC 4329a},
and \object{NGC 4151}, and
$\theta\sim67\degr$ for \object{MCG 8-11-11}.
These large inclination angles reflect the small amount of reflection in these objects. Using Fig.~\ref{rdemu}, one can check that these angles are in good agreement with the previously derived $R$ parameter values~: 1.03 ($\tetref=60\degr$), 1.64 ($\tetref=0\degr$) and 0.43 ($\tetref=65\degr$), respectively for \object{IC 4329a}, \object{MCG 8-11-11} and \object{NGC 4151}.

 However, our \chisq  values are worse because the \ion{Fe}{} line is not a free parameter in our model and our absorption model is rough. Ignoring the data bins in the 6.-7. range (model $c$ and $d$) improves the fits significantly. This means that our predicted \ion{Fe}{} line does not reproduce the data very well. An analysis of the residuals shows that the observed lines are stronger and broader in these three objects. Such large equivalent widths are usually interpreted as iron overabundance. The broadening could arise from Doppler and general relativistic effects. A secondary component produced elsewhere could also be present.
\begin{table*}[!]
\begin{flushleft}
\caption{\label{results}\textit{Best fit parameters and $\chi^{2}$ for models (a)  $\beta$ frozen, (b) $\beta$ free parameter, (c) $\beta$ frozen, 6.-7.~keV channels ignored, (d) $\beta$ free, 6.-7.~keV channels ignored. $E_\mathrm{c}$ is given in keV, $N_\mathrm{H}$ in $10^{22}\,cm^{-2}$.  }}

\begin{tabular}{lcccccc}

\hline
 & $\Gamma$ & $E_\mathrm{c}$ & $\beta$ & $\mu$ & $N_\mathrm{H}$ & $\chi^{2}/\nu$\\
\hline\\

\multicolumn{2}{l}{\object{IC 4329a}}\\
\hline 
(a)& $1.95^{+0.03}_{-0.01}$ & $284.^{+120}_{-60}$ & 1.0 (f)    & $0.16_{-0.01}^{+0.02}$ & $0.46^{+0.02}_{-0.02}$ & 129/104 \\
\hline
(b) & $1.87^{+0.06}_{-0.11}$ & $112._{-111}^{+240}$ & $0.39^{+0.30}_{-0.13}$ & $0.18^{+0.01}_{-0.01}$ & $0.46^{+0.02}_{-0.02}$ & 127/103 \\
\hline
(c) &  $1.96^{+0.01}_{-0.02}$ & $306.^{+136}_{-60}$ & 1.0 (f)   & $0.16 ^{+0.02}_{-0.02}$ & $0.46^{-0.02}_{-0.02}$ & 118/101\\
\hline
(d) & $1.97^{+0.01}_{-0.10}$ & $210.^{+112}_{-210}$ & $2.6^{+6.1}_{-2.2}$ & $0.15^{+0.03}_{-0.04}$  & $0.47^{+0.02}_{-0.03}$ & 118/100\\
\hline\\

\multicolumn{2}{l}{\object{MCG 8-11-11}}\\
\hline
 (a) & $1.71^{+0.04}_{-0.04}$ & $201.^{+96}_{-55}$ & 1.0 (f) & $0.38^{+0.05}_{-0.07}$ & $0.27^{+0.02}_{-0.02}$ & 913/884\\
\hline
 (b) & $1.71^{+0.04}_{-0.13}$ & $203.^{+110}_{-202}$ & $0.96^{+1.8}_{-0.6}$ & $0.38^{+0.06}_{-0.07}$ & $0.27^{+0.02}_{-0.02}$ & 913/883\\
\hline
 (c) & $1.71^{+0.05}_{-0.05}$ & $217.^{+64}_{-64}$ & 1.0 (f) & $0.36^{+0.13}_{-0.13}$ & $0.27^{+0.02}_{-0.02}$ & 814/827\\
\hline
 (d) & $1.66^{+0.01}_{-0.01}$ & $116.^{+242}_{-71}$ & $0.51 ^{+2.9}_{-0.12}$ & $0.41^{+0.10}_{-0.20}$ & $0.27^{+0.02}_{-0.02}$ & 814/826\\
\hline\\

\multicolumn{2}{l}{\object{NGC 4151}}\\
\hline
 (a) & $1.80^{+0.01}_{-0.01}$ & $235.^{+25}_{-19}$ & 1.0 (f) &  $0.15^{+0.01}_{-0.01}$ & $9.4 ^{+0.1}_{-0.1}$ & 137/99\\
\hline
 (b) & $1.87^{+0.04}_{-0.03}$ & $211.^{+20}_{-20}$  & $2.4 ^{+0.6}_{-0.5}$ & $0.16^{+0.03}_{-0.02}$ & $9.8^{+0.2}_{-0.2}$ & 100/98\\
\hline
 (c) & $1.84^{+0.01}_{-0.08}$ & $282.^{+14}_{-78}$ & 1.0 (f) &  $0.18^{+0.01}_{-0.05}$ & $9.5^{+0.1}_{-0.4}$ & 131/96\\
\hline
 (d) & $1.87^{+0.06}_{-0.03}$ & $210.^{+25}_{-15}$ & $2.4^{+1.2}_{-0.5}$ & $0.16^{+0.05}_{-0.02}$ & $9.8^{+0.2}_{-0.3}$ & 93/95\\
\hline
\end{tabular}

\end{flushleft}

\end{table*}

Actually, a detailed modelling of the \ion{Fe}{} line would require more assumptions than we did, making it difficult a self consistent calculation in the framework of the anisotropic illumination model.
However, we can see by eyes in Fig.~\ref{data} that our estimates of the line are not in complete disagreement with the data.

In \object{IC 4329a} and \object{MCG 8-11-11} the data do not constrain the detailed shape of the cut-off. So adding the \be parameter (model $b$ and $d$) does not improve significantly the fit. And as \be and \ec are strongly correlated, similar $\chi^{2}$ values  can be achieved with either small or large value of both \be and \ec. In \object{MCG 8-11-11} the best fit values of \be in model $b$ is found to be close to 1., while ignoring the 6-7 keV channels lead to $\be=0.51$ but with a $\chi^{2}$ value very close to that obtained with \be fixed to 1 (model $c$). For \object{IC 4329a} we found two local minima corresponding to $\be\sim0.4$ , $\ec\sim110keV$ and $\be\sim2.7$, $\ec\sim210keV$ with a slight difference in  $\chi^{2}$ values. The first minimum is the best fit in model (b), while the second becomes the best fit when 6-7 keV channels are ignored.
  On the other hand, in \object{NGC 4151}, the OSSE data have a higher photon statistics which enables to better constrain the model. The cut-off is clearly sharper than exponential which had led Zdziarski et al. (\cite{zdziarski}) to interpret it as a thermal one. Our nonthermal model gives a good description of the spectrum provided that $\be>1$, with a best fit value of $\be$ close to 2.4.           

   
\section{Discussion}


In the previous section we found that the AIM is consitent with the observed X-ray/$\gamma$-ray spectra from \object{IC 4329a}, \object{MCG 8-11-11} and \object{NGC 4151}, provided that the inclination angle is nearly edge-on. This would require that, at least in these three Seyfert 1 galaxies, the dusty torus is inexistent or the geometry of the obscuring material enables us to see directly their nuclei at large inclination angles. To our knowledge, such an hypothesis can not be ruled out by present observations. There are also some hints that the inclination is larger than usually thought in some unobscured Seyferts. 

In \object{NGC 4151}, HST observations of the ionisation cone in the extended narrow line region require a  larger opening angle and an inclination $>60\degr$ (Evans et al. \cite{evans}). The large absorption column density and the radio jet in \object{NGC 4151} are also suggestive of an edge-on Seyfert. This led Cassidy \& Raine (\cite{cassidy}) to propose a modified version of the unified model based on detailed analysis of \object{NGC 4151}. In this scheme, the observed ionisation cone is due to the geometry of the scatterer rather than to collimation by the torus.The absorption of Seyfert 2 spectra occurs in a flared accretion disc, so the opening angle of the torus is not required to be small. For \object{NGC 4151} they derive an inclination of $58\degr$.
 
In the case of \object{IC 4329a}, a polarimetric study (Wolstencroft et al. \cite{wolstencroft}), assuming that the polarization arises from scattering on the inner edge of the optically thick torus, shows that \object{IC 4329a} is unlikely to be seen pole-on. Rather, it suggests that our line of sight goes through the edge of the torus so that the central nucleus is partially obscured.
This would be consistent with a large inclination if the torus is 
geometrically thin and, on the other hand, it would also explain the 
large column density obtained in the best fits --almost 10 times the galactic 
column density (see also Walter\&Fink \cite{walter}). We can equally remark that a large reddening has been observed in the optical and ultraviolet (Marziani 
et al. \cite{marziani92}) and that \object{IC 4329a} is one of the rare almost edge-on Seyfert 1 
galaxies (Keel \cite{keel80}).

We are thus possibly dealing with three highly inclined systems.

However, our results are clearly different from detailed studies of the 
\ion{Fe}{K} line profiles observed by ASCA (Nandra et al. \cite{nandra97}).
They found broad asymmetric redshifted lines consistent with emission in the inner regions of an accretion disc rotating around a black hole. The relativistic effects on the line profile are strongly angle dependent (Fabian et al. \cite{fabian89}; Laor \cite{laor}). They derive a mean inclination of only  $30\degr$ for their sample of Seyfert 1s. For \object{NGC 4151} and \object{IC 4329a}, they found viewing angles respectively in the range $9\degr-33\degr$ and $10\degr-26\degr$ degrees. At least for \object{MCG 6-30-15} and \object{NGC 4151}, it seems difficult to find a plausible  alternative explanation (Fabian et al. \cite{fabian95}) such as Compton downscattering.                            
Note however that the modelling of the continuum can affect the line profile. On the other hand, the used fitting model requires had hoc assumptions on the inner and outer radii of the disc and on the disc emissivity profile and assumes that all of the fluorescence arise from the disc. Moreover, this model also predicts small inclination angles for a sample of Seyfert 2s and NELG (Turner et al. \cite{turner}) in contradiction with the unified model.

A more problematic issue is the prediction of reflection 
dominated Seyfert 1s for pole-on inclination angles. However, a  proeminent reflection component should have strong effects below ten keV and thus should be detected by GINGA and ASCA. Indeed, these objects would present a very intense \ion{Fe}{}line with equivalent widths of 1-2 keV and a very hard spectrum in this band ($\Gamma\sim1$). 
 It is interesting to note that observations of the X-ray background requires discrete sources with photon index in the 2-10 keV range $\Gamma\sim1.5$ (Ueda et al. \cite{Ued98}). Although these sources are merely Seyfert 2 galaxies (Sakano et al. \cite{sak98}), the X-ray background would be consistent with reflection dominated Seyfert 1s ( Fabian et al. \cite{fabianbkg}).

A few reflection dominated Seyferts have effectively been reported (Reynolds et al. \cite{reynolds}, Matt et al. \cite{matt}, Malaguti et al. \cite{malaguti}), but all of them are Seyfert 2s. The reflection component is then interpreted as reflection on the inner edge of the torus, the nuclear emission being strongly absorbed.

In the GINGA sample of 27 Seyfert 1s and NELG (Nandra \& Pound \cite{nandra94}), the largest R values are 3-4. This means that, according to Fig.~\ref{rdemu}, none of them would have an inclination lower than  $\sim50\degr$. Unless there is a strong selection effect against low inclination system , this situation is most unlikely. The data do not suggest the existence of a selection bias which would be moreover difficult to explain. The AIM in its present form is thus ruled out.
\begin{figure}

 \resizebox{\hsize}{!}{\includegraphics{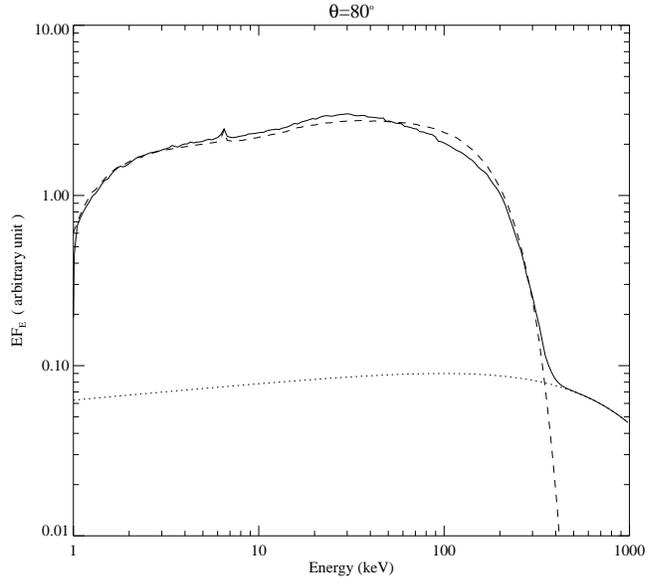}}
 \caption{Spectrum obtained for a disk fully ionised in its external parts (solid line). The half-angle substended by the neutral region is 45 \degr. The parameters are $\mu=0.98$, $\al=1.9$, $\nh=0.47$ and $\ec=3$~MeV. The direct emmission is shown in dotted line. As a comparison the dashed line shows the overall spectrum obtained with best fit parameters of model $d$ for \object{IC4329a}}.  
 \label{ionise}
\end{figure}

However, some improvements of the AIM could make it more consistent with the observations. First,
  the source could be closer to the disc. Then, the solid angle subtended by the reflecting matter is reduced due to the central hole in the disk. Yet, it appears that the
  reduction of the hump while significant is not sufficient to solve our
  problem, unless the source is very close to the black hole. For an
  inclination of $10\degr$, we found that the source has to be lower
  than a tenth of the radius of the hole. This is rather unrealistic.
  However, we do not account for relativistic effects in these estimates
  and it is clear that, at such distances of the black hole, they become
  very important. Introduction of relativistic effects is deferred to a
  future work.

 On the other hand, a particularly relevant improvement would be to consider the ionisation of the disk upper layers due to illumination.
The ionisation degree of the reflecting material depends on the incident flux and  the density.
We made preliminary estimates of the density profile at the surface of the disk. It turns out to be denser close to the black hole. So that paradoxically, it should be neutral in its internal parts and highly ionized at greater radii.

In the ionised regions the reflected component would arise with very few photelectric absorption mimicking the primary spectrum at low energy, leading to an apparent reduced reflection hump.
As a first estimate of this effect we calculated the spectra predicted by the AIM assuming that the central region is neutral, while the external region is fully ionised (no photoabsorption, reflection on free electrons) . The border between the two zones is parametrized by $\phi$ the half-angle subtended by the neutral region as seen from the source.
Fig.~\ref{ionise} compares the spectrum obtained in our fit of \object{IC4329a} (model $c$) and a spectrum obtained with $\phi= 45 \degr$, and an inclination of $10 \degr$. The spectral shapes are very similar. We can equally remark that, as the direct emission is negligible for small inclinations, the cut-off energy is no longer constrained to be at a few hundred keV (Ec=3 MeV in Fig~\ref{ionise} ). This could have interesting consequences for non-thermal models.

\section{Conclusion}

We calculated the reflection component predicted by the AIM. It is found to be stronger than in isotropic model and it dominates the high energy spectrum for small inclination angles. Such a strong reflection component is unobserved in Seyfert 1s. So, to be consistent with the data, the AIM requires that all the observed Seyferts are seen with a large viewing angle. Obtaining acceptable fits of IC4329a, MCG 8-11-11 and NGC4151 spectra requires inclinations of 70 \degr - 80 \degr . While it is possible that these three individual objects are very inclined, it seems very unlikely to be the case for all the Seyfert 1s galaxies observed by GINGA, ASCA and BeppoSAX. So we consider that the AIM, in its present form, is ruled out. However, we showed that it could be reconciled with the data if one admit that a fair part of the disk is strongly ionised. A detailed study of the effect of ionisation as well as other improvements of the model that could reduce the reflection hump are under works.

\begin{acknowledgements}
We are very grateful to C.~Done, E.J.~Grove, W.N.~Johnson,
 G.~Madejski, D.~Smith  and particularly to P.~Grandi, P.~Magdziarz, and A.A.~Zdziarski for providing us with the data. 
\end{acknowledgements}

\end{document}